\documentclass[aps,prb,twocolumn,showpacs,letterpaper,longbibliography,superscriptaddress]{revtex4-1}

\usepackage[colorlinks=true, citecolor=blue, linkcolor=blue, urlcolor=blue]{hyperref}
\usepackage{graphicx,dcolumn,longtable,epsfig}
\usepackage[usenames]{color}
\usepackage{amssymb}
\usepackage{amsmath}
\usepackage{bm}
\usepackage{footnote}
\usepackage{float}
\usepackage{subfigure}
\usepackage{color}  

\usepackage{ulem}
\usepackage[T1]{fontenc}

\newcommand{\be}{\begin{equation}}
\newcommand{\ee}{\end{equation}}

\def\bea{\begin{eqnarray}}
\def\eea{\end{eqnarray}}

\begin{document}
\title{Robustness of bipolaronic superconductivity to electron-density-phonon coupling}

\author{Chao Zhang}
\email{chaozhang@ahnu.edu.cn}
\affiliation{Department of Physics, Anhui Normal University, Wuhu, Anhui 241000, China}

\begin{abstract}
We study bipolaron formation and bipolaronic superconductivity on a square lattice, where electrons couple to both local Holstein phonons via on-site charge density and nonlocal bond Su-Schrieffer-Heeger phonons via modulation of hopping amplitudes. Using an unbiased Diagrammatic Monte Carlo method, we investigate how the interplay between these two types of electron-phonon coupling affects the bipolaron binding energy, effective mass, spatial extent (quantified by the mean-squared radius), and the superconducting transition temperature $T_c$. We find that, in some parameter space, the moderate Holstein coupling, though detrimental to $T_c$ when acting alone, can enhance superconductivity when combined with the bond SSH coupling by further compressing the bipolaron without significantly increasing its mass. Similarly, introducing bond SSH coupling into a Holstein bipolaron reduces its size while keeping the effective mass nearly unchanged, leading a higher $T_c$. These effects give rise to nonmonotonic behavior and reveal a cooperative regime in which both couplings work together to enhance superconductivity. We further examine phonon frequency asymmetry, particularly the case $\omega_H/t = 2\omega_B/t$, and show that in the deep adiabatic regime, adding Holstein coupling can even raise $T_c$ when combined with bond SSH coupling. These results highlight the distinct and complementary roles of local Holstein and non-local bond SSH electron-phonon couplings, and suggest strategies for optimizing high-$T_c$ superconductivity in systems with multiple phonon modes.
\end{abstract}

\maketitle


\section{Introduction}
\label{sec:sec1}

Bipolaron formation, where two electrons bind via lattice distortions, provides a plausible route to superconductivity in strongly electron-phonon coupled systems. In low-density or polaronic regimes, such composite bosons can condense into a coherent quantum state at finite temperature. Unlike conventional Bardeen–Cooper–Schrieffer (BCS) pairing, this mechanism does not rely on a Fermi surface and is particularly relevant when electron-phonon coupling dominate. However, the resulting superconducting transition temperature $T_c$ is highly sensitive to the bipolaron’s effective mass and spatial extent, which in turn depends on the nature of the coupling~\cite{PhysRevB.111.184513, PhysRevB.108.L220502, zhangarXiv, PhysRevX.13.011010}.

Two representative types of electron-phonon coupling have been extensively studied. The Holstein type interaction couples electronic charge density to local lattice distortions, typically producing tightly bound but heavy bipolarons with suppressed $T_c$~\cite{ Chakraverty, PhysRevLett.84.3153, PhysRevB.69.245111,Alexandrov:1999fy}. In contrast, the bond Su-Schrieffer-Heeger (SSH) type interaction couples bond phonons to electron hopping amplitudes, generating lighter and more delocalized bipolarons that can sustain higher $T_c$~\cite{PhysRevB.111.184513, PhysRevB.108.L220502, zhangarXiv, PhysRevX.13.011010}.

In realistic materials, electron-phonon interactions are often more complex than those described by a single coupling mechanism. Both local (Holstein-type) and non-local (bond SSH-type) couplings can coexist due to the presence of multiple phonon branches with different phonon frequency symmetries. Local distortions typically modulate on-site electronic energies, while bond phonons affect the hopping amplitudes between neighboring sites. The interplay between these coupling types may significantly influence polaron and bipolaron formation, and hence the emergence and stability of superconductivity. Understanding how these interactions compete or cooperate is therefore essential for capturing the full behavior of electron-phonon systems.

Despite extensive studies of the Holstein model~\cite{Holstein2000,Bonca2000,Kornilovitch1998} and the bond-type Su-Schrieffer-Heeger (SSH) model~\cite{PhysRevB.104.035143,PhysRevLett.121.247001,PhysRevB.108.075156,PhysRevB.109.L220502} in isolation, the combined effects of these two electron-phonon coupling mechanisms remain less understood. In Ref.~\cite{PhysRevB.109.165119}, the authors investigated the properties of a single polaron subject to both Holstein and bond SSH couplings. They found a non-monotonic dependence of the effective mass on the bond coupling strength $g_B$, particularly when the Holstein coupling $g_H$ lies within the strong-coupling regime. This interplay suggests that it may be possible to achieve both a light polaron effective mass and strong binding—favorable conditions for bipolaron formation and superconductivity.

In this work, we investigate a two-electron system on a square lattice coupled to both Holstein and bond-type SSH phonons using the unbiased Diagrammatic Monte Carlo (DiagMC) method. We systematically study how the interplay between local and nonlocal electron-phonon interactions shapes the bipolaron properties—binding energy, effective mass, and spatial size—and determine the corresponding superconducting transition temperature $T_c$. Our results reveal a cooperative regime in which weak-to-moderate Holstein and bond SSH couplings can jointly enhance bipolaron compactness and mobility, thus promoting high-temperature superconductivity.

Our results reveal a cooperative regime where moderate Holstein and bond SSH couplings can enhance $T_c$. Starting from an optimal bond SSH bipolaron, weak Holstein coupling further compresses its size without significantly increasing its mass, leading to a nonmonotonic enhancement in $T_c$. Similarly, introducing bond SSH coupling to a Holstein bipolaron reduces its size and but keeps the effective mass nearly unchanged, leading to an increasing of $T_c$. We also examine the role of phonon frequency asymmetry and find that high-frequency Holstein modes mitigate mass enhancement, while low-frequency phonon modes can even further boost $T_c$ when combined with bond SSH coupling. These findings demonstrate the distinct yet complementary roles of different electron-phonon couplings, and point to a viable strategy for enhancing bipolaronic superconductivity through multimode electron-phonon coupling. 

The remainder of this paper is organized as follows. In Sec.~\ref{sec:sec2}, we introduce the model and outline the physical observables computed using the DiagMC method. In Sec.~\ref{sec:sec3} A, we present results for the case $\omega_H/t = \omega_B/t$ at various phonon frequencies $\omega_B/t = 1.0,\ 0.5,\ 0.3$, and $0.2$. Section~\ref{sec:sec3} B discusses the asymmetric case $\omega_H = 2\omega_B$, which reflects more realistic cases. Finally, in Sec.~\ref{sec:sec4}, we summarize our main conclusions.

\section{Model and Method}
\label{sec:sec2}

We study bipolaron formation in a strongly correlated, dilute two-electron system on a two-dimensional square lattice. The electrons interact with both local Holstein-type and nonlocal bond SSH-type phonons. The total Hamiltonian is given by
\begin{equation}
H = H_e + H_{\text{ph}} + H_{\text{int}},
\end{equation}
where the electronic part is
\begin{equation}
H_e = -t \sum_{\langle ij \rangle, \sigma} \left( c_{j,\sigma}^\dagger c_{i,\sigma} + \text{H.c.} \right) + U \sum_i n_{i,\uparrow} n_{i,\downarrow},
\end{equation}
with $c_{i,\sigma}^\dagger$ ($c_{i,\sigma}$)
creating (annihilating) an electron of spin $\sigma \in \{ \uparrow, \downarrow \}$ at site $i$, and $n_{i,\sigma} = c_{i,\sigma}^\dagger c_{i,\sigma}$. We set the hopping amplitude $t = 1$ as the unit of energy throughout this work. 
The on-site Hubbard repulsion is fixed at $U/t = 8$, following Ref.~\cite{PhysRevX.13.011010}, where this value was shown to yield the optimal superconducting transition temperature $T_c$.

The phonon Hamiltonian includes both on-site and bond phonon modes:
\begin{equation}
H_{\text{ph}} = \omega_H \sum_i \left( b_i^\dagger b_i + \frac{1}{2} \right)+ \omega_B \sum_{\langle ij \rangle} \left( b_{\langle ij \rangle}^\dagger b_{\langle ij \rangle} + \frac{1}{2} \right),
\end{equation}
where $b_i$ and $b_{\langle ij \rangle}$ are bosonic annihilation operators associated with local Holstein type and non-local bond type phonons, respectively. The phonon frequencies are denoted by $\omega_H$ (Holstein) and $\omega_B$ (bond SSH).

Electron-phonon couplings are given by
\begin{equation}
H_{\text{int}} = g_H \sum_i n_i X_i + g_B \sum_{\langle ij \rangle, \sigma} \left( c_{j,\sigma}^\dagger c_{i,\sigma} + \text{H.c.} \right) X_{\langle ij \rangle},
\end{equation}
with the phonon displacement operators defined as $X_i = b_i + b_i^\dagger$ and $X_{\langle ij \rangle} = b_{\langle ij \rangle} + b_{\langle ij \rangle}^\dagger$. The coupling strengths $g_H$ and $g_B$ control the Holstein and bond SSH couplings, respectively.

To characterize the bipolaron properties, we employ the diagrammatic Monte Carlo (DiagMC) method, which is well established in the literature~\cite{PhysRevB.105.L020501, PhysRevB.107.L121109, PhysRevB.62.6317, PhysRevLett.81.2514, PhysRevLett.123.076601, PhysRevB.106.L041117, Prokofev:2008jz,Prokofev:2008it} and has been successfully applied to study both Holstein and bond SSH polaronic systems. Specifically, we compute the bipolaron binding energy, effective mass, and mean-squared radius as functions of the coupling parameters.

For generic a few body system with Hamiltonian $\hat{H}$, imaginary-time evolution projects an arbitrary state onto the ground state. Although the number of electrons in the system is small, the inclusion of an infinite number of phonon degrees of freedom renders it a genuine many-body problem. In this method, we measure the standard $n$-particle Green's Function: $G_{ba}(\tau)$. The decay of the propagator $G_{ba}(\tau)$ at long $\tau$ encodes the ground-state energy $E_g$:
\begin{equation}
G_{ba}(\tau) \xrightarrow[\tau \to \infty]{} \langle b | g \rangle \langle g | a \rangle e^{-\tau E_g}.
\end{equation}
where $|a\rangle$ and $|b\rangle$ are states with nonzero overlap with the ground state $|g\rangle$.

By fitting this exponential decay, we extract $E_g$ for both single-polaron and bipolaron sectors. The binding energy is then defined as
\begin{equation}
\Delta_{\text{BP}}(\mathbf{k}) = 2E_{\text{p}}(\mathbf{k}) - E_{\text{BP}}(\mathbf{k}),
\end{equation}
where $E_{\text{p}}(\mathbf{k})$ and $E_{\text{BP}}(\mathbf{k})$ are the single-polaron and bipolaron energies at momentum $\mathbf{k}$.

To evaluate the polaron effective mass $m_{\text{BP}}^*$, we analyze the spatial broadening of the imaginary-time propagator. For large $\tau$ and relative displacement $\mathbf{R}$ between initial and final center-of-mass coordinates~\cite{PhysRevLett.74.2288}, the Green’s function scales as
\begin{equation}
G_{ba}(\tau, \mathbf{R}) \sim \frac{A_{ba} e^{-\tau E_g}}{\tau^{d/2}} \exp\left(-\frac{m_{\text{BP}}^*R^2}{2 \tau}\right),
\end{equation}
with $A_{ba}$ is the coefficient.

This leads to the estimator
\begin{equation}
\overline{\mathbf{R}^2}(\tau) = \frac{\sum_{ab} W_{ab} G_{ba}(\tau, \mathbf{R}) \mathbf{R}^2 }{\sum_{ab} W_{ab} G_{ba}(\tau, \mathbf{R})} \xrightarrow[\tau \to \infty]{} \frac{d}{m_{\text{BP}}^*} \tau,
\end{equation}
with $W_{ab}$ the weight of the states $| a \rangle$ and $| b \rangle$.
The effective mass is then extracted from the slope of $\overline{\mathbf{R}^2}(\tau)$ in the large-$\tau$ limit.

The mean-squared radius of the bipolaron provides a measure of its spatial extent. It is obtained by sampling the interparticle distance $\mathbf{r}$ at the midpoint of long bipolaron trajectories:
\begin{equation}
R^2_{\text{BP}} = \sum_{\mathbf{r}} \left(\frac{r}{2}\right)^2 P(\mathbf{r}),
\end{equation}
where $P(\mathbf{r})$ is the probability distribution for finding the two electrons at separation $\mathbf{r}$ in the ground state.

In the dilute limit, tightly bound bipolarons behave as composite bosons and can undergo a two-dimensional Berezinskii-Kosterlitz-Thouless (BKT) transition. The corresponding superfluid transition temperature can be approximated by~\cite{PhysRevB.104.L201109, PhysRevLett.130.236001, PhysRevLett.87.270402, PhysRevLett.100.140405}:
\begin{equation}
T_c \approx
\begin{cases}
\displaystyle \frac{C}{m^*_{\mathrm{BP}} R^2_{\mathrm{BP}}}, & \text{if } R^2_{\mathrm{BP}} \gtrsim 1 \\
\displaystyle \frac{C}{m^*_{\mathrm{BP}}}, & \text{otherwise}
\end{cases}
\end{equation}
where $m^*_{\mathrm{BP}}$ is the bipolaron effective mass, and $C \approx 0.5$ is a prefactor appropriate for dilute 2D Bose gases. This formulation provides a reliable upper bound for $T_c$, even when weak boson-boson interactions are present.

All simulations are carried out on a two-dimensional square lattice of size $L = 140$ with open boundary conditions. Benchmark tests confirm that finite-size effects are negligible for the computed observables at this system size.

In a previous study, we investigated the case of pure bond SSH coupling (with no Holstein coupling, so $g_H/t = 0$) and examined bipolaron formation across several phonon frequencies $\omega_B/t = 1.0,\ 0.5,\ 0.3,\ 0.2$, identifying regimes where light, compact bipolarons emerge and the superconducting transition temperature $T_c$ is maximized~\cite{PhysRevX.13.011010}. In the present work, we extend this analysis by introducing a finite Holstein coupling ($g_H/t > 0$) to study how the addition of on-site density-type interactions affects bipolaron properties. This enables us to assess the cooperative or competing roles of local Holstein and nonlocal bond SSH couplings, and to determine how their interplay influences bipolaron binding energy, effective mass, mean-squared radius, and the resulting superconducting transition temperature $T_c$.



\section{Interplay between Holstein and bond SSH Couplings}
\label{sec:sec3}

In this section, we investigate the interplay between Holstein and bond SSH electron-phonon couplings by considering two representative phonon frequency ratios: (A) $\omega_H/t = \omega_B/t$ and (B) $\omega_H/t = 2\omega_B/t$. The latter case is motivated by realistic materials, where multiple phonon branches typically have distinct energy scales. Throughout this work, we fix the on-site Hubbard interaction at $U/t = 8.0$, which was previously identified as an optimal value for maximizing the bipolaronic transition temperature in the bond SSH model~\cite{PhysRevX.13.011010}.

For each case, we examine two complementary scenarios.

\textbf{(i) Adding Holstein coupling to the bond SSH model:} Starting from the bond SSH model at the previously identified optimal $T_c/\omega_B$ point (as reported in Ref.~\cite{PhysRevX.13.011010}) for various phonon frequencies $\omega_B/t = 1.0,\ 0.5,\ 0.3,\ 0.2$, we gradually introduce Holstein coupling $g_H/t$. Prior studies have shown that bond SSH coupling can produce lightweight, compact bipolarons and relatively high $T_c$, whereas Holstein coupling tends to generate heavier, more spatially extended bipolarons, typically suppressing $T_c$. Here, we investigate how the addition of Holstein coupling modifies the bipolaron properties and alters the superconducting transition temperature $T_c$ previously associated with the pure bond SSH model. 

\textbf{(ii) Adding bond SSH coupling to the Holstein model:} Here, we start with the pure Holstein model and the optimal coupling $g_H/t= 1.414$ for $\omega_H/t = 0.5$ was reported in Ref.~\cite{PhysRevX.13.011010}. At this intermediate coupling, we gradually introduce bond SSH coupling $g_B/t$ to examine how the nonlocal phonon channel affect properties of bipolaron, like the effective mass and the mean-squared radius, thereby affect $T_c$.

These two complementary pathways allow us to systematically explore the cooperative and competing effects of coexisting local Holstein and nonlocal bond SSH type electron-phonon couplings across a broad range of phonon frequencies and coupling strengths.

\begin{figure}[t]
\includegraphics[width=0.5\textwidth]{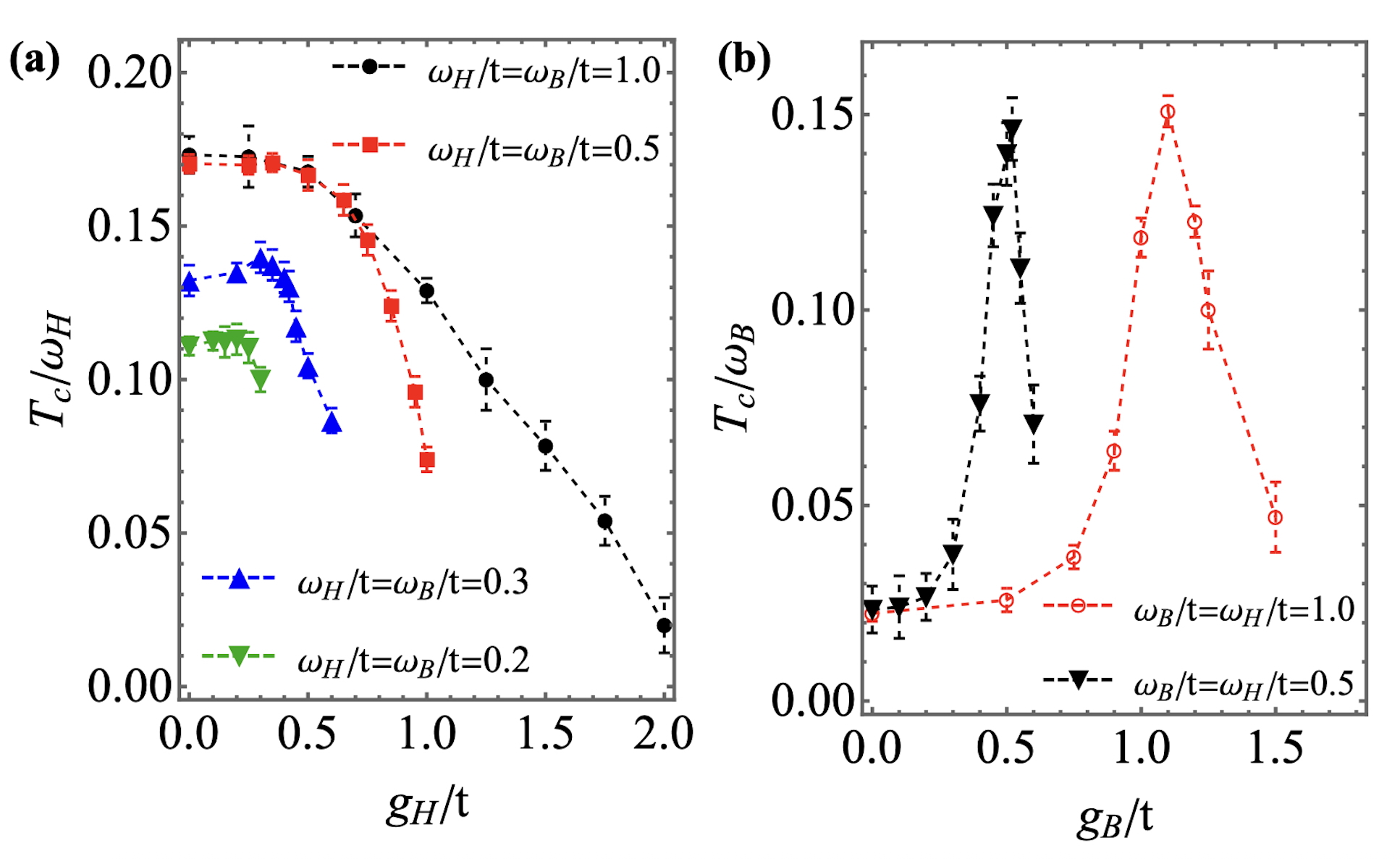}
\caption{
(a) Bipolaronic superconducting transition temperature $T_c$ (in units of $\omega_H$) as a function of Holstein coupling $g_H/t$ for various phonon frequencies, with $\omega_H/t = \omega_B/t$ and bond SSH coupling $g_B/t$ fixed at the optimal values identified for the pure bond SSH model~\cite{PhysRevX.13.011010}: $g_B/t = 1.25$ for $\omega_B/t = 1.0$ (black dots), $0.75$ for $0.5$ (red squares), $0.5477$ for $0.3$ (blue upward triangles), and $0.433$ for $0.2$ (green downward triangles).  (b) $T_c$ (in units of $\omega_B$) as a function of bond SSH coupling $g_B/t$ for fixed Holstein coupling $g_H/t = 1.25$ and $\omega_H/t = \omega_B/t$. Results are shown for $\omega_B/t = 1.0$ (black downward triangles) and $0.5$ (red circles). The addition of bond SSH coupling significantly enhances $T_c$, revealing the cooperative role of nonlocal bond SSH electron-phonon coupling in Holstein bipolarons.
}
\label{Figure1}
\end{figure}

\subsection{ $\omega_H/t$= $\omega_B/t$}

Figure~\ref{Figure1}(a) shows the bipolaronic superconducting transition temperature $T_c$ (in units of the Holstein phonon frequency $\omega_H$) as a function of Holstein coupling strength $g_H/t$, for various phonon frequencies where $\omega_H/t = \omega_B/t$. The bond SSH coupling $g_B/t$ is fixed at values corresponding to the optimal $T_c$ in the pure bond SSH type electron-phonon coupling as reported in Ref.~\cite{PhysRevX.13.011010}: $g_B/t = 1.25$ for $\omega_B/t = 1.0$ (black dots), $g_B/t=0.75$ for $\omega_B/t=0.5$ (red squares), $g_B/t=0.5477$ for $\omega_B/t=0.3$ (blue upward triangles), and $g_B/t=0.433$ for $\omega_B/t=0.2$ (green downward triangles). As $g_H/t$ increases from zero, $T_c/\omega$ remains nearly unchanged or slightly increases in the weak-coupling regime, particularly for $\omega_H/t = \omega_B/t = 0.3$ and $0.2$. However, upon further increasing $g_H/t$, $T_c$ eventually decreases. The suppression of $T_c$ becomes more abrupt for lower phonon frequencies, reflecting the stronger mass renormalization associated with Holstein coupling in the deep adiabatic regime. Figure~\ref{Figure1}(b) presents $T_c$ (in units
of the bond SSH phonon frequency $\omega_B$) as a function of bond SSH coupling $g_B/t$ for fixed Holstein coupling $g_H/t = 1.25$ and equal phonon frequencies $\omega_H/t = \omega_B/t$. The black downward triangles correspond to $\omega_B/t = 1.0$, and the red dots to $\omega_B/t=0.5$. In both cases, increasing the bond SSH coupling in a Holstein bipolaron leads to a significant enhancement in $T_c$, demonstrating that the nonlocal bond SSH coupling can counteract the adverse effects of Holstein coupling and effectively restore or even boost superconductivity.

\begin{figure}[t]
\includegraphics[width=0.5\textwidth]{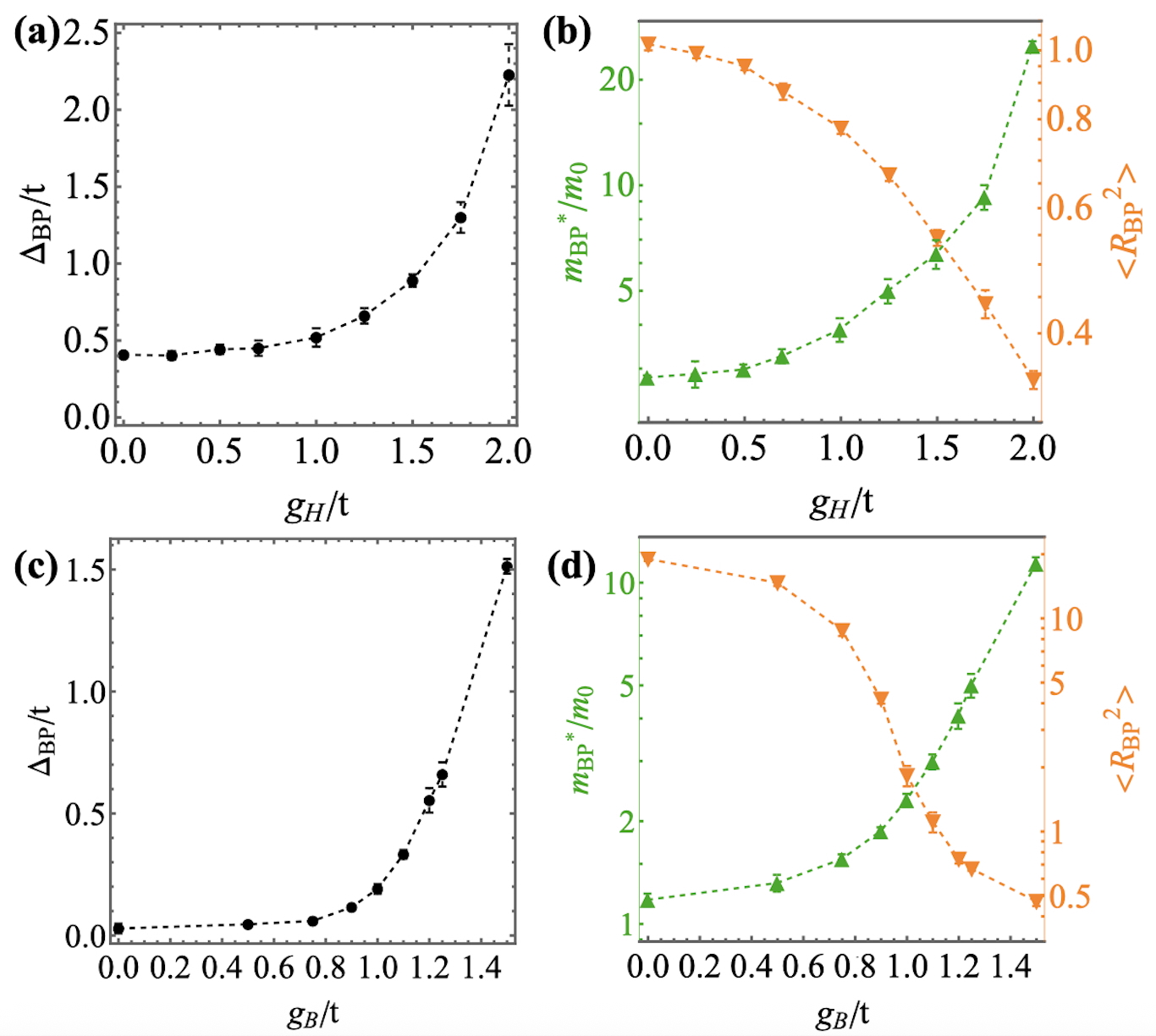}
\caption{
Bipolaron properties at equal adiabaticity $\omega_B/t = \omega_H/t = 1.0$. 
(a) Binding energy $\Delta_{\text{BP}}$ (in units of $t$), and 
(b) effective mass $m^*_{\text{BP}}/m_0$ (green upward triangles, with $m_0 = 2m_e = 1/t$) and mean-squared radius $R^2_{\text{BP}}$ (orange downward triangles), plotted as functions of Holstein coupling $g_H/t$ at fixed bond SSH coupling $g_B/t = 1.25$. 
(c,d) Same quantities as in (a,b), now shown as functions of bond SSH coupling $g_B/t$ at fixed Holstein coupling $g_H/t = 1.25$ and $\omega_H/t=1.0$. 
Error bars, if not visible, are smaller than the symbol size.
}
\label{Figure2}
\end{figure}

To better understand the microscopic origin of the superconducting trends, we analyze bipolaron properties—binding energy $\Delta_{\text{BP}}/t$, effective mass $m^*_{\text{BP}}/m_0$ with $m_0=1/t$, the bare mass of two electrons, and mean squared radius $R^2_{\text{BP}}$—as functions of $g_H/t$ (case (i)) and $g_B/t$ (case (ii)), as shown in Fig.~\ref{Figure2}. All results are obtained from QMC simulations at fixed phonon frequencies $\omega_H/t = \omega_B/t = 1.0$. 

Figures~\ref{Figure2}(a) and \ref{Figure2}(b) show the dependence on Holstein coupling $g_H/t$, with bond SSH coupling fixed at $g_B/t = 1.25$, which corresponds to the optimal $T_c/\omega$ in the pure bond SSH model for $\omega_B/t=1.0$. As $g_H/t$ increases, the binding energy $\Delta_{\text{BP}}/t$ increases and the bipolaron radius $R_{\text{BP}}^2$ decreases, indicating that the electrons are pulled closer together by the additional Holstein coupling. The effective mass $m^*_{\text{BP}}/m_0$ grows slowly at first but begins to rise more steeply beyond $g_H/t \sim 0.7$, reflecting the increasing Holstein type coupling-induced mass renormalization. This heavier mass arises from stronger local lattice dressing, which hinders the bipolaron’s ability to move coherently. This competition between size reduction and mass enhancement explains the nonmonotonic $T_c/\omega_H$ behavior in Fig.~\ref{Figure1}(a). At small $g_H/t$, the bipolaron becomes more compact without a large increase in mass, keeping $T_c/\omega_H$ stable or slightly enhanced due to increased pairing strength and retained mobility. However, as $g_H/t$ becomes larger, the rapid growth of the effective mass outweighs the benefit of size reduction, leading to a suppressed $T_c/\omega_H$. This result highlights that compactness alone is insufficient for superconductivity—bipolarons must remain light to achieve high $T_c$.

In contrast, Figs.\ref{Figure2}(c) and \ref{Figure2}(d) show how bond SSH coupling influences bipolaron properties at fixed Holstein coupling $g_H/t = 1.25$, with $\omega_B/t = \omega_H/t = 1.0$. As $g_B/t$ increases, the bipolaron binding energy $\Delta_{\text{BP}}/t$ increases, the effective mass $m^*_{\text{BP}}/m_0$ also grows, and the mean squared radius $R^2_{\text{BP}}$ decreases. Interestingly, when starting from this moderate Holstein coupling $g_H/t = 1.25$—a regime where $T_c$ is relatively low—the system hosts a loosely bound bipolaron with a large spatial extent ($R^2_{\text{BP}} \sim 20$) and only slightly enhanced mass ($m^*_{\text{BP}}/m_0 \gtrsim 1$). In this case, the low transition temperature originates mainly from the delocalized nature of the bipolaron rather than mass renormalization. Introducing bond SSH coupling into this state leads to a notable increase in $T_c$, as the nonlocal interaction efficiently compresses the bipolaron without greatly enhancing its mass. However, at larger $g_B/t$, the effective mass rises more significantly due to stronger lattice dressing, eventually offsetting the gain from localization and resulting in a suppressed $T_c$.

In summary, our results reveal a cooperative regime in which local Holstein and nonlocal bond SSH electron-phonon couplings act synergistically to enhance bipolaronic superconductivity. When a moderate Holstein coupling is added to the bond SSH model, the local coupling compresses the bipolaron, while the bond coupling maintains a relatively low effective mass. This combination produces small and mobile bipolarons, leading to an enhanced superconducting transition temperature $T_c$. However, as the Holstein coupling becomes stronger, the associated mass renormalization dominates, reducing carrier mobility and suppressing superconductivity. This nonmonotonic evolution of $T_c$ highlights the competition between spatial confinement and mass enhancement. On the other hand, when the bond SSH coupling is introduced into a Holstein-dominated system, it efficiently reduces the bipolaron size and mitigates the heavy-mass problem inherent to the local coupling. At moderate bond coupling strengths, $T_c$ increases markedly due to stronger pairing to form a compact bipolaron. Yet, beyond a certain $g_B/t$, the rapid increase of effective mass again leads to a downturn of $T_c$. Together, these results demonstrate that tuning the balance between local Holstein and bond SSH electron-phonon couplings can optimize bipolaron compactness and mobility, thereby stabilizing superconductivity at higher transition temperatures.

\begin{figure}[t]
\includegraphics[width=0.48\textwidth]{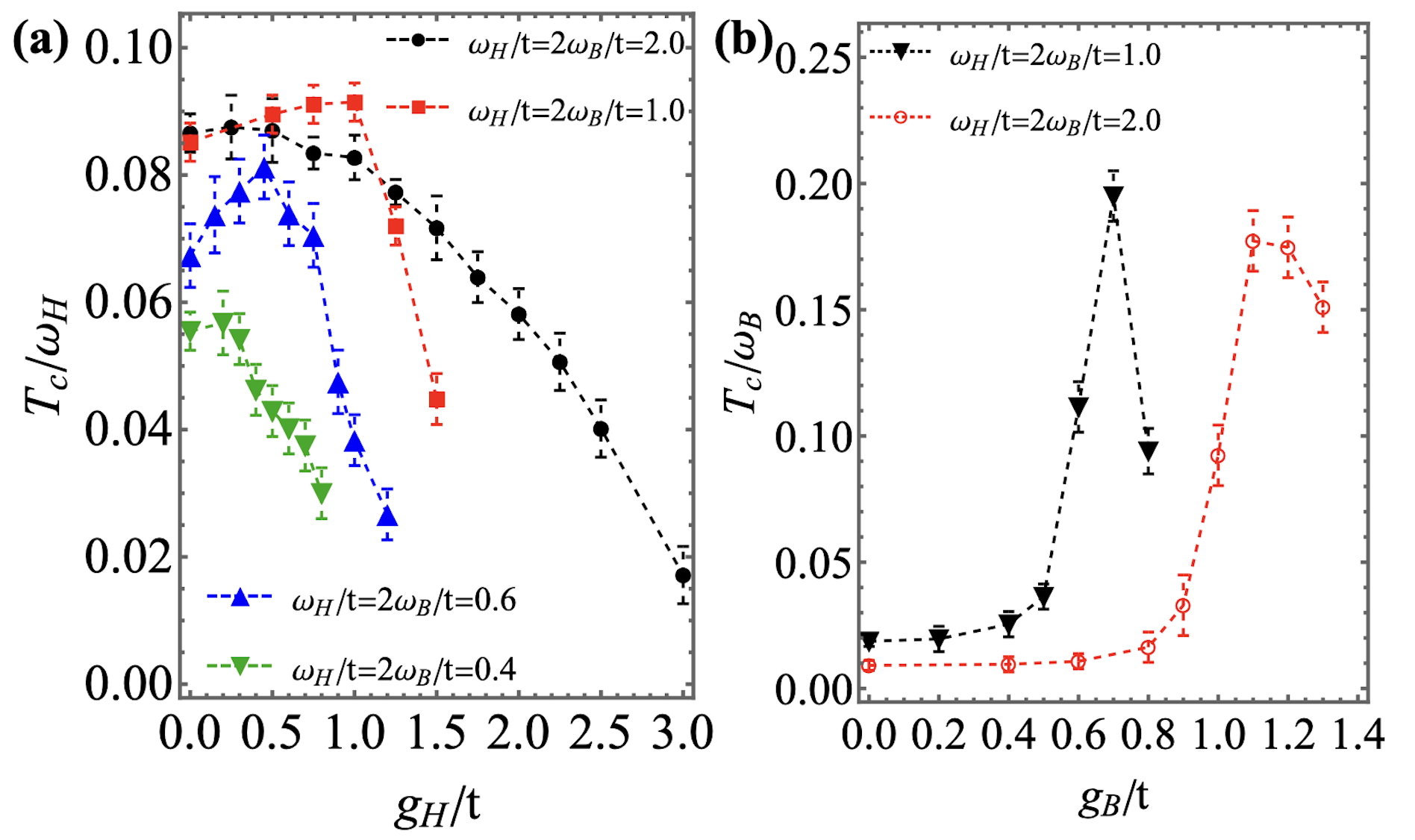}
\caption{(a) Bipolaronic superconducting transition temperature $T_c$ (in units of $\omega_H$) as a function of Holstein coupling $g_H/t$ for various phonon frequencies, with $\omega_H/t = 2\omega_B/t$. The bond SSH coupling $g_B/t$ is fixed at the optimal values identified for the corresponding pure bond model: $g_B/t = 1.25$ for $\omega_B/t = 1.0$ (black dots), $0.75$ for $0.5$ (red squares), $0.5477$ for $0.3$ (blue upward triangles), and $0.433$ for $0.2$ (green downward triangles). (b) $T_c$ (in units of $\omega_B$) as a function of bond SSH coupling $g_B/t$ for fixed Holstein coupling $g_H/t = 1.25$ and $\omega_H/t = 2\omega_B/t$. Results are shown for $\omega_B/t = 1.0$ (black downward triangles) and $0.5$ (red circles). The addition of bond SSH coupling significantly enhances $T_c$, revealing the cooperative role of nonlocal electron-phonon couplings in Holstein bipolarons.
}
\label{Figure3}
\end{figure}

\subsection{$\omega_H/t = 2\omega_B/t$}

To better emulate realistic materials with multiple phonon branches, we investigate the case where the Holstein phonon frequency is twice that of the bond SSH mode: $\omega_H/t = 2\omega_B/t$. This hierarchy reflects physical scenarios where high-energy local optical phonons coexist with lower-frequency bond SSH phonons.

\begin{figure}[t]
\includegraphics[width=0.48\textwidth]{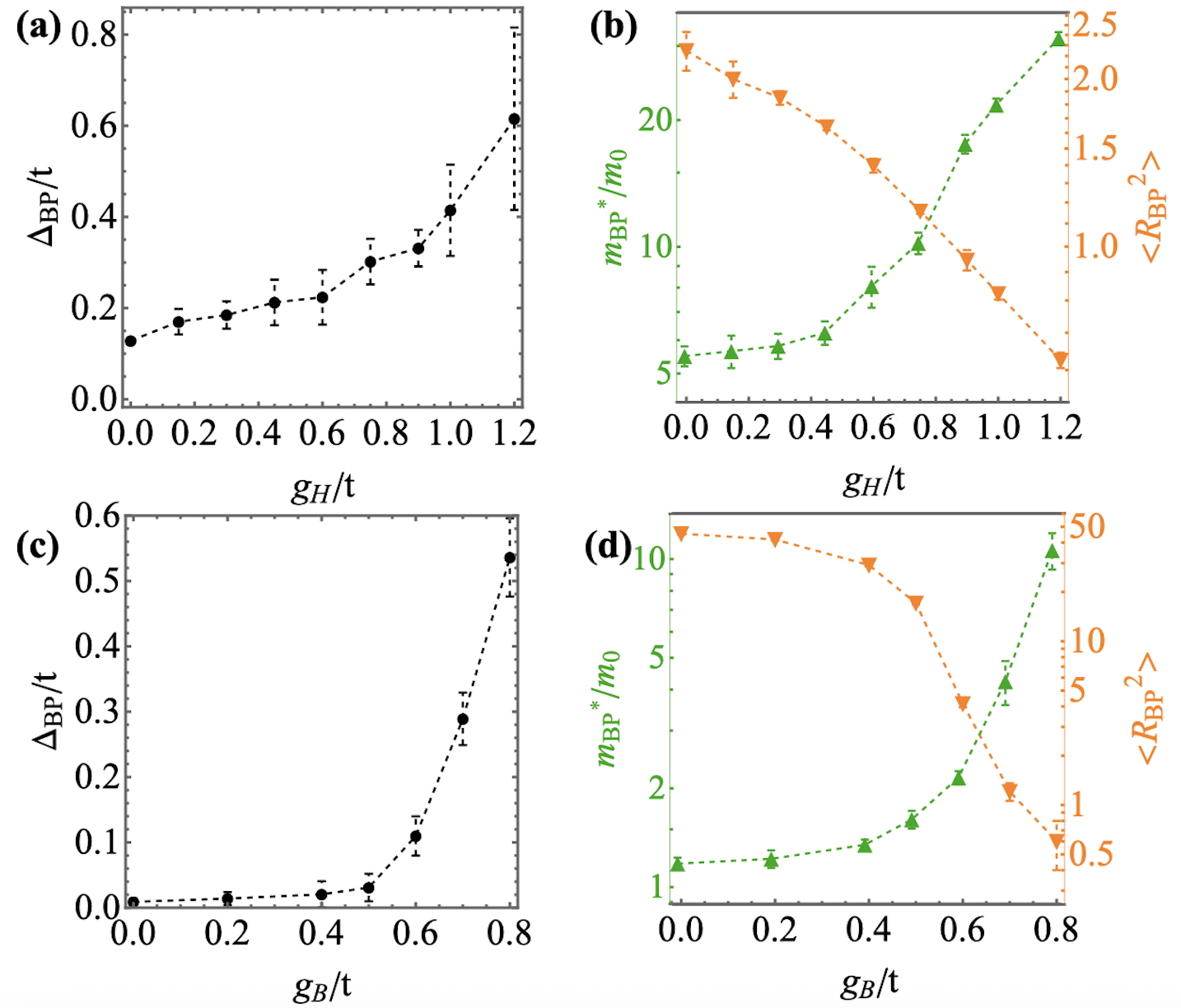}
\caption{Bipolaron properties at unequal adiabaticity $\omega_H/t = 2\omega_B/t$. (a) Binding energy $\Delta_{\text{BP}}$ (in units of $t$), and  
(b) effective mass $m^*_{\text{BP}}/m_0$ (green upward triangles, with $m_0 = 2m_e = 1/t$) and mean-squared radius $R^2_{\text{BP}}$ (orange downward triangles), plotted as functions of Holstein coupling $g_H/t$ at fixed bond SSH coupling $g_B/t = 0.5477$, for $\omega_H/t = 0.6$ and $\omega_B/t = 0.3$. (c,d) Same quantities as in (a,b), now shown as functions of Holstein coupling $g_H/t$ at fixed bond SSH coupling $g_B/t = 1.25$, for $\omega_H/t = 1.0$ and $\omega_B/t = 0.5$. Error bars, if not visible, are smaller than the symbol size.
}
\label{Figure4}
\end{figure}

Figure~\ref{Figure3}(a) shows the superconducting transition temperature $T_c$ as a function of Holstein coupling $g_H/t$ for several phonon frequencies under the asymmetric condition $\omega_H/t = 2\omega_B/t$. As in the equal-frequency case, the bond SSH coupling $g_B/t$ is fixed at the optimal values previously identified for the pure bond SSH model~\cite{PhysRevX.13.011010}. For all values of $\omega_B/t$, $T_c$ exhibits a nonmonotonic dependence on $g_H/t$: it initially increases slightly, reaches a broad maximum, and then gradually decreases. Compared to the symmetric condition ($\omega_H/t = \omega_B/t$), both the enhancement and suppression are more moderate. We also explore the complementary scenario in which the Holstein coupling is fixed at $g_H/t = 1.25$, and bond SSH coupling $g_B/t$ is varied for two representative cases: $\omega_H/t = 2\omega_B/t = 1.0$ and $2.0$ as shown in Fig.~\ref{Figure3}(b). In both cases, introducing a small $g_B/t$ enhances $T_c$ by compressing the bipolaron and preserving its mobility. However, as $g_B/t$ increases beyond a threshold, the effective mass grows rapidly, leading to a subsequent decline in $T_c$. This behavior qualitatively mirrors the trends observed in the symmetric case. These results demonstrate that phonon frequency asymmetry offers an additional tuning knob for optimizing bipolaronic superconductivity by balancing the bipolaron effective mass and bipolaron size.

This trend is supported by the bipolaron properties shown in Fig.~\ref{Figure4}. Panels (a)–(b) correspond to $\omega_H/t = 0.6$, $\omega_B/t = 0.3$, with fixed $g_B/t = 0.5477$. As $g_H/t$ increases, the binding energy $\Delta_{\text{BP}}$ rises slightly, while both the effective mass $m^*_{\text{BP}}$ and the radius $R_{\text{BP}}^2$ change modestly. Panels (c)–(d) show results for $\omega_H/t = 1.0$ and $\omega_B/t = 0.5$, with fixed $g_B/t = 1.25$, exhibiting a similar trend: the bipolaron energy and effective mass evolve smoothly with $g_H/t$, but the radius $R_{\text{BP}}^2$ drops rapidly, in contrast to the nearly symmetric case ($\omega_H/t = \omega_B/t$).

These findings suggest that when the Holstein mode is higher in frequency than the bond SSH mode, the superconducting temperature $T_c$ still remains robust over a wider range of parameters. This frequency asymmetry effectively broadens the cooperative regime, offering a design principle for optimizing bipolaronic superconductivity.

\section{Conclusion}
\label{sec:sec4}
In this work, we investigate bipolaron formation and bipolaronic superconductivity in a two-dimensional square lattice where electrons couple to both local Holstein and nonlocal bond SSH phonon modes. Using unbiased diagrammatic Monte Carlo simulations, we systematically explore how the interplay between Holstein-type and bond SSH-type electron-phonon couplings affects bipolaron properties—including binding energy, effective mass, mean-squared radius—as well as the superconducting transition temperature $T_c$.

We find that while Holstein coupling alone tends to suppress superconductivity due to strong mass renormalization, its addition to an bond SSH bipolaron can enhance $T_c$ in the weak to intermediate regime by compressing the bipolaron without significantly increasing its mass. On the other hand, introducing bond SSH coupling to a Holstein bipolaron reduces its spatial extent and preserves mobility, leading to substantial improvements in $T_c$. These findings reveal a cooperative regime in which moderate couplings of both types act synergistically to optimize superconducting properties. 

Moreover, we show that phonon frequency asymmetry plays a key role: when the Holstein phonon has a higher frequency than the bond SSH phonon, its detrimental impact on effective mass is mitigated, further broadening the cooperative window. These results highlight the importance of multimode electron-phonon coupling and suggest a microscopic strategy for engineering light, compact bipolarons conducive to high-$T_c$ superconductivity.

\begin{acknowledgments}
We thank Nikolay Prokof'ev, Boris Svistunov, John Sous, David Reichman, Mona Berciu, and Andrew Millis for inspire this work and helpful discussions. 

\end{acknowledgments}

\bibliography{bond}

\end{document}